# Observation of charged excitons in hole-doped carbon nanotubes using photoluminescence and absorption spectroscopy

Ryusuke Matsunaga[1], Kazunari Matsuda[1], and Yoshihiko Kanemitsu[1,2]

[1]*Institute for Chemical Research, Kyoto University, Uji, Kyoto 611-0011, Japan*

[2]*Photonics and Electronics Science and Engineering Center, Kyoto University, Kyoto 615-8510, Japan*

## Abstract

**We report the first observation of trions (charged excitons), three-particle bound states consisting of one electron and two holes, in hole-doped carbon nanotubes at room temperature. When *p*-type dopants are added to carbon nanotube solutions, the photoluminescence and absorption peaks of the trions appear far below the $E_{11}$ bright exciton peak, regardless of the dopant species. The unexpectedly large energy separation between the bright excitons and the trions is attributed to the strong electron-hole exchange interaction in carbon nanotubes.**

Because of their unique mechanical, optical, and electronic properties, single-walled carbon nanotubes have been of great interest in broad research fields over the past decade [1]. One of the most prominent characteristics of carbon nanotubes is the strong correlation between carriers due to the quantum confinement that occurs in one-dimensional (1D) structures of about 1 nm diameter. Electron-electron repulsive and electron-hole attractive interactions play important roles in the electronic and optical properties [2], resulting in the formation of excitons with huge binding energies in semiconducting carbon nanotubes [3], and even in metallic carbon nanotubes [4].

Despite the importance of carrier correlations in carbon nanotubes, many-particle bound states, such as exciton-exciton or exciton-electron(hole) complexes, are not yet fully understood. Excitonic molecules (biexcitons) have not been observed in carbon nanotubes even under intense excitation conditions at low temperatures [5]. In addition, there has been no report of observation of exciton-carrier bound states (charged excitons or trions) in carbon nanotubes, although many optical measurements have been performed with carrier-doping [6-11]. Meanwhile, the carrier-doping effect on nanotube field-effect transistors has been investigated for controlling and switching the optical properties of nanotube devices [12,13]. Additionally, a recent *ab initio* calculation has shown that the band gap can be tuned by dynamical screening of acoustic plasmons in hole-doped carbon nanotubes [14]. Thus, experimental studies of the optical properties of carrier-doped carbon nanotubes are quite important not only for understanding many-particle correlations in 1D structures, but also for future nanoscale device applications.

In this Letter, we report the existence of a new excited state below the lowest ($E_{11}$) singlet exciton state in hole-doped single-walled carbon nanotubes. In both photoluminescence (PL) and absorption spectroscopy, a new peak appears with hole doping at room temperature, regardless of the dopant species. This PL peak is completely different from the luminescence of optically forbidden excitons reported recently [15-19]. Moreover, the energy separation between the new state and the $E_{11}$ bright exciton state shows clear "family patterns," similar to the exciton binding energy in carbon nanotubes. These results indicate that the new peak exhibits formation of trions. The extremely large energy separation (100-200 meV) between the bright excitons and the trions is attributed to large singlet-triplet exciton splitting due to the short-range Coulomb interaction in carbon nanotubes.

We prepared various kinds of carbon nanotube samples and dopants to check whether the spectral change due to hole doping originated from a specific dopant type. Carbon nanotubes synthesized by the HiPCO and CoMoCAT methods were dispersed in toluene solutions with 1 wt% poly[9,9-dioctylfluorenyl-2,7-diyl] (PFO), 60 min of moderate bath sonication, 24 h of vigorous sonication with a tip-type sonicator, and ultracentrifugation at 30 000 $g$ for 20 min [20]. We denoted these samples as HiPCO-PFO and CoMoCAT-PFO nanotubes, respectively. For hole doping, we prepared a $p$-type dopant, 2,3,5,6-tetrafluoro-7,7,8,8-tetracyanoquinodimethane ($F_4$TCNQ), in a 1 mg/mL solution of toluene [10]. We also prepared another type of nanotube sample: the carbon nanotubes synthesized by the HiPCO method were dispersed in $D_2O$ solutions with 0.1 wt% sodium dodecyl sulfonate (SDS), 60 min of moderate bath sonication, 60 min of vigorous sonication with a tip-type sonicator, and ultracentrifugation at 138 000 $g$ for 7 h [21]. We refer to these as HiPCO-SDS nanotubes. The ultracentrifugation time for the HiPCO-SDS nanotubes was relatively long to exclude as many residual bundles as possible. Because the bundles are not susceptible to hole doping and hinder the doping-induced spectral changes, the samples including very few bundles are important for this study. For comparison with $F_4TCNQ$, we prepared another *p*-type dopant, 4-amino-1,1-azobenzene-3,4-disulphonic acid (AB) in a $10^{-2}$ mol/L solution in $D_2O$ [10]. We also prepared a hydrochloric acid (HCl) solution (pH 2) for altering the pH, because $H^+$ ions adsorbed on the surface of nanotubes act as *p*-type dopants [6-9]. All optical experiments were carried out at room temperature.

Figure 1(a) shows a contour map of PL excitation (PLE) for nondoped CoMoCAT-PFO nanotubes, which included many (7,5) nanotubes and a small amount of (7,6) nanotubes. The dotted circles in Fig. 1(a) indicate the PL of the $E_{11}$ bright excitons in the (7,5) and (7,6) nanotubes. The side peak indicated by the arrow in Fig. 1(a) is the phonon sideband of the *K*-momentum dark excitons in (7,5) carbon nanotubes [17,18]. Figure 1(b) shows the PLE map after a *p*-type dopant $F_4TCNQ$ (200 µg/mL) was added to the CoMoCAT-PFO nanotube solutions. The PL intensity of the $E_{11}$ bright excitons in Fig. 1(b) was reduced by a factor of about 50, compared with that of the nondoped carbon nanotubes in Fig. 1(a), due to the hole doping [10]. At the lower-energy side, new PL peaks appeared, as indicated by the solid circles in Fig. 1(b). These PL peaks are clearly associated with (7,5) and (7,6) carbon nanotubes. Figure 1(c) shows the normalized PL spectra of the nondoped and doped CoMoCAT-PFO nanotubes [Fig. 1(a) and (b), respectively] at the excitation photon energy of 1.9 eV. The doping-induced new PL peaks are indicated by the arrows in Fig. 1(c). The relative PL intensities of the bright excitons for (7,6), (7,5), and (6,5) nanotubes were changed by the doping. This is because the PL intensities of larger-diameter tubes are more sensitive to the hole doping [10]. The full widths at half maximum of the (7,5) bright excitons and the new PL peak at 1.0 eV were about 22 ± 2 and 36 ± 3 meV, respectively, and almost unchanged by the dopant concentration.

The new peaks were also observed by absorption spectroscopy with hole doping. Figure 2(a) shows the absorption spectra of CoMoCAT-PFO nanotubes with $F_4TCNQ$ solutions added. The inset shows an enlarged view of the low-energy side. With increasing $F_4TCNQ$ concentration, the new absorption peaks appeared as indicated by the arrows A and B in Fig. 2(a). Compared with the new PL peaks observed in Fig. 1(a), we found that the absorption peaks A and B were associated with (7,5) and (7,6) nanotubes, respectively. Figure 2(b) shows the normalized absorption intensity of the $E_{11}$ bright exciton peak (left axis) and the new peak (right axis) of (7,5) carbon nanotubes as a function of $F_4TCNQ$ concentration. While the absorption intensity of the $E_{11}$ exciton decreased due to the hole doping [10], the absorbance of the new state increased.

We also performed hole doping to HiPCO-PFO nanotubes by adding $F_4TCNQ$ and observed the appearance of the same new PL and absorption peaks, corresponding to each type of chirality (see supplementary material [22]).

If the new state originates from any chemical complex between $F_4TCNQ$ and carbon nanotubes, the new peak should not appear with other dopants, or at least the energy position of the new peak should depend on the reduction potential of the dopant. To examine this, we performed hole doping using another *p*-type dopant, the AB solution. Figure 3(a) shows the PLE map of nondoped HiPCO-SDS nanotubes. The white dotted circles in Fig. 3(a) indicate the $E_{11}$ bright exciton PL peaks. Figure 3(b) shows the PLE map after the AB solution was added. Figure 3(b) is divided into three parts by the white dotted lines: the lower-left, middle, and upper-right parts correspond to the data with adding AB solutions of 50, 100, and 200 μL, respectively. This is because the spectral change due to AB solutions was drastic and larger-diameter tubes [corresponding to the lower-left part in Fig. 3(b)] were more susceptible to the doping effect than smaller-diameter tubes [10]. In Fig. 3(b), the PL of the $E_{11}$ bright excitons almost disappeared due to hole doping, except for some small-diameter tubes. Then, as indicated by the solid circles in Fig. 3(b), new low-energy PL peaks were observed at the low-energy sides of each nanotube chiral type. Although the reduction potential of the *p*-type dopant AB was different from that of $F_4TCNQ$ [10], the new peak appeared as in the case of $F_4TCNQ$. Moreover, we confirmed that the same new PL peak was also observed when HCl solutions (*p*-type dopant, $H^+$) were added instead of AB solutions (see supplementary material [22]). These results indicate that the new PL peak observed in hole-doped carbon nanotubes is independent of the dopant species, and that the new state exhibits intrinsic properties of hole-doped carbon nanotubes, rather than any chemical complex with a specific dopant.

Figure 4 shows the tube-diameter dependence of the energy separation Δ between the bright exciton state and the new state observed in the PL and absorption spectra. The energy separation depends strongly on the tube diameter, but not on the species of dopants. The value of Δ is about 130 meV for 1-nm diameter. This diameter dependence of the new peak is completely different from the phonon sideband of the *K*-momentum dark excitons [17,18] or the luminescence of triplet excitons [18,19]. The new PL and absorption peaks appear with very small Stokes shifts. In Fig. 4, we also show the value of $2n+m$ for each chirality with the HiPCO-SDS data, where (*n*,*m*) is the chiral index. The HiPCO-SDS data show clear family patterns, which are well-known optical characteristics of carbon nanotubes, similar to the exciton binding energy [23]. This result confirmed that the new state arises from the intrinsic properties of carrier-doped carbon nanotubes, rather than specific dopants.

Here, we consider the origin of the intrinsic low-energy PL and absorption peaks. Recent theoretical calculations have shown that stable trions can exist in carbon nanotubes [24]. A trion binding energy is defined as the energy required for dissociating a trion into a free hole and an exciton. The calculated trion binding energy in carbon nanotubes for dielectric constant of 3.5 is about $40/d$ meV [24], where *d* is the tube diameter (nm), and is much larger than that in other II-VI [25] or III-V [26] compound semiconductors. It means that trions in carbon nanotubes are detectable even at room temperature. Thus, we assign the new state below the lowest excitons in hole-doped carbon nanotubes to trions, which have not been experimentally reported in carbon nanotubes thus far.

The energy separation between the bright excitons and trions observed in our experiments for the 1-nm-diameter tube was about 130 meV, which is surprisingly large.

We attribute the extremely large energy separation to the electron-hole exchange interaction originating from the short-range Coulomb interaction in carbon nanotubes [27], which has not been considered in the previous trion calculations [24]. For conventional semiconductors where the trions have been observed, the exchange interaction is very weak and the singlet-triplet exciton splitting has not been considered. However, the excitons in carbon nanotubes have large singlet-triplet splitting as much as several tens of meV, as calculated from theories [27-29] and recently observed in experiments [18,19]. This is because the strong exchange interaction between an electron and a hole in carbon nanotubes significantly lifts up the energy level of the singlet bright exciton state [27]. Thus, the trion binding energy corresponds to the energy separation between the trion states and the lowest triplet exciton states, and the singlet-triplet exciton splitting cannot be neglected for the observation of trions in carbon nanotubes (inset of Fig. 4). The recent calculation of trions including the short-range Coulomb interaction also supports our results that the trion states are below the triplet exciton states [30]. The singlet-triplet exciton splitting has been reported as about $70/d^2$ meV [18]. Assuming the trion binding energy of $\sim 60/d$ meV, which is close to the theoretical calculation [24], we show the energy separation $\Delta = 70/d^2 + 60/d$ as the solid curve in Fig.4. Although the coefficients depend on the dielectric constant, the solid curve is in good agreement with the experimental data. Thus, we conclude that the new peaks in hole-doped carbon nanotubes exhibited trion formation at room temperature.

In the past, there has been a lack of discussion of trions in systems with strong exchange interactions, such as organic molecules. The formation of room-temperature trions in hole-doped carbon nanotubes will stimulate more detailed experiments in the wide range of temperature and increase our understanding of many-particle correlations in such low-dimensional materials. Furthermore, recent theoretical studies have proposed novel schemes to optically manipulate the carrier spins in carbon nanotubes through trion states [31, 32]. Even though the spin-orbit interaction is very weak in carbon nanotubes, both the selective spin preparation and the transfer of the spin information to a photon will become possible in combination with an external magnetic field and an optical cavity [31]. The existence and thermal stability of the trions in nanotubes will facilitate the realization of the optical spin manipulation in carbon nanotubes.

In conclusion, we reported the first observation of trions in carbon nanotubes by adding *p*-type dopants at room temperature. The trions in carbon nanotubes have an unexpectedly large energy separation (100-200 meV) from the bright excitons. We attributed this to the stronger exchange interaction in carbon nanotubes compared with conventional semiconductors. Our findings provide further insights into electron correlation and exchange effects in carbon nanotubes.

We thank Y. Miyata, K. Watanabe, and K. Asano for helpful discussions and M. Murata, Y. Murata, T. Umeyama, and H. Imahori for experimental equipments. This study was supported by KAKENHI (Grants No. 20340075, No. 20048004, and No. 20104006) and JSPS (Grant No. 21-1890).

**Figure captions**

**Fig. 1** (Color online) PLE contour maps of (a) nondoped CoMoCAT-PFO nanotubes, (b) CoMoCAT-PFO nanotubes doped with $F_4$TCNQ solutions (200 μg/mL). The dotted circles show the $E_{11}$ bright excitons and the solid circles show the new PL peaks due to hole doping. (c) The normalized PL spectra of the nondoped and doped CoMoCAT-PFO nanotubes [from (a) and (b), respectively] at the excitation photon energy of 1.9 eV. The doping-induced new PL peaks of (7,5) and (7,6) nanotubes are indicated by the solid and open arrows, respectively.

**Fig. 2** (Color online) (a) The hole-doping dependence of the absorption spectra of CoMoCAT-PFO nanotubes. The inset shows an enlarged view of the lower-energy region where the new peaks appear. The new peaks indicated by the arrows A and B are associated with (7,5) and (7,6) nanotubes, respectively. (b) The hole-doping dependence of the absorption intensity of the $E_{11}$ bright excitons (left axis) and the new peak (right axis) of (7,5) carbon nanotubes.

**Fig. 3** (Color online) (a) PLE contour map of nondoped HiPCO-SDS nanotubes. The intensity scale in the right part (> 1.35 eV) of (a) is increased tenfold for clarity. The dotted circles show the $E_{11}$ bright excitons. (b) PLE contour map of hole-doped HiPCO-SDS nanotubes. The lower-left, middle, and upper-right regions correspond to hole doping with AB solutions of 50, 100, and 200 μL, respectively. The solid circles show the new low-energy PL peaks observed with hole doping.

**Fig. 4** (Color online) Tube diameter $d$ dependence of the energy separation $\Delta$ between the bright exciton state and the new state observed in PL and absorption spectra. The green numbers show the value of *2n+m* for (*n,m*) nanotubes. The calculated solid curve shows the sum of the exchange splitting of $70/d^2$ meV and the trion binding energy of $60/d$ meV. Inset shows the schematic of the energy levels of the singlet and triplet excitons ($X_S$ and $X_T$) and the trions ($X^+$). A free hole is denoted as $h^+$.

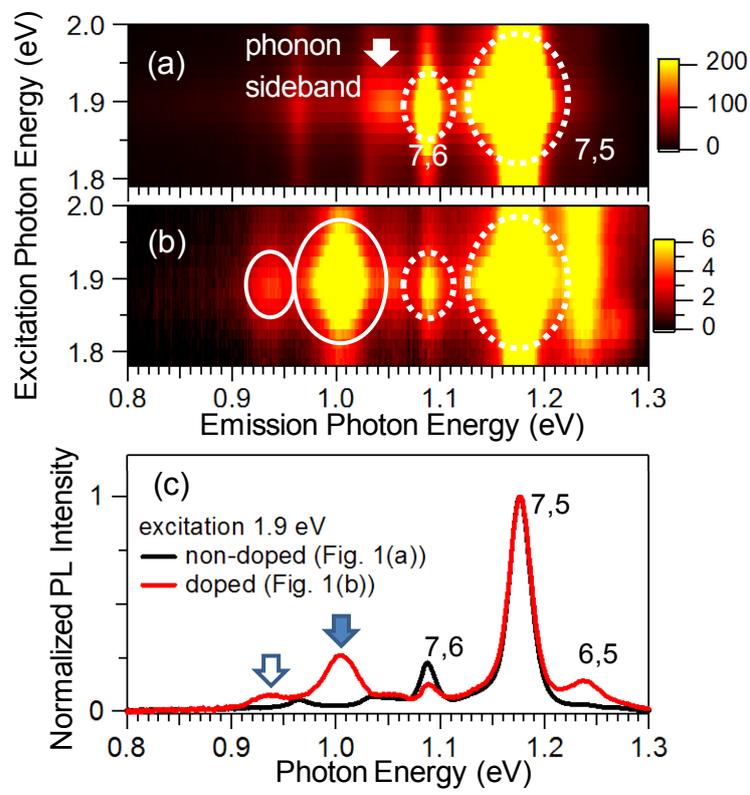

Fig. 1 Matsunaga et al.,

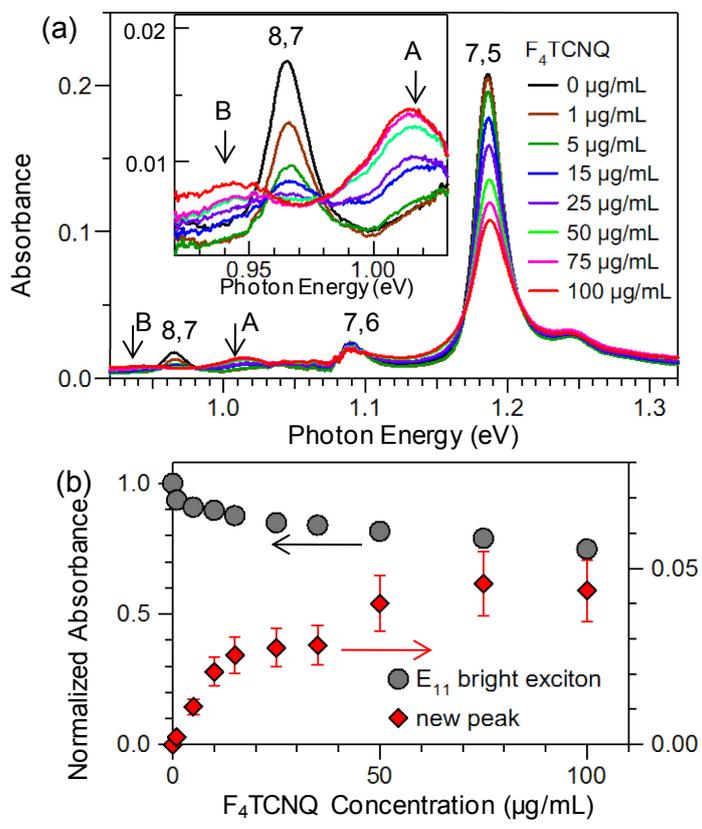

Fig. 2 Matsunaga et al.,

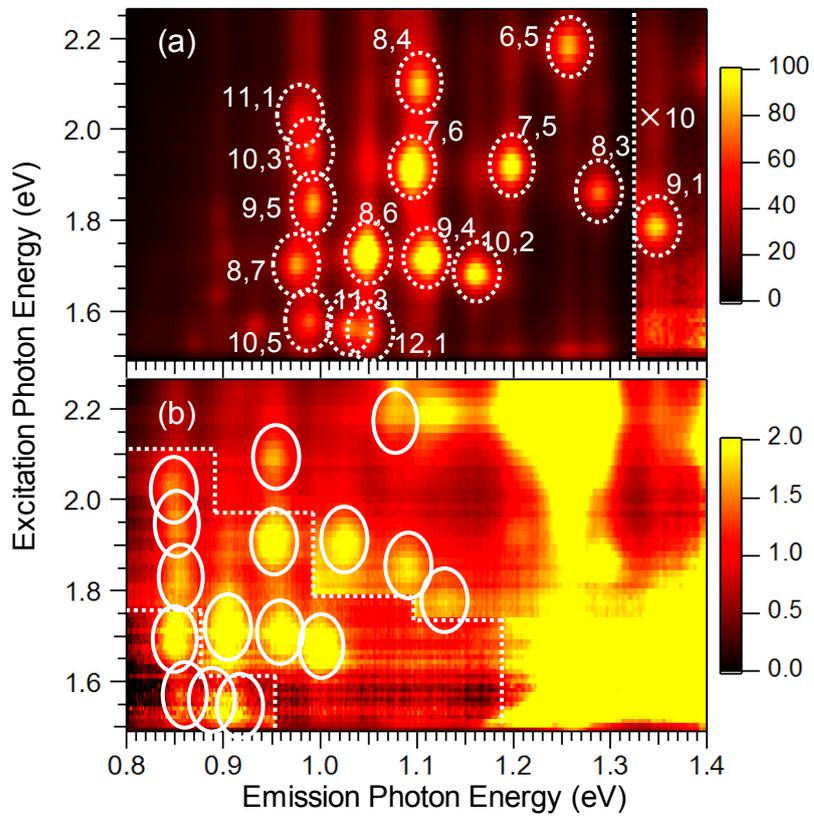

Fig. 3 Matsunaga et al.,

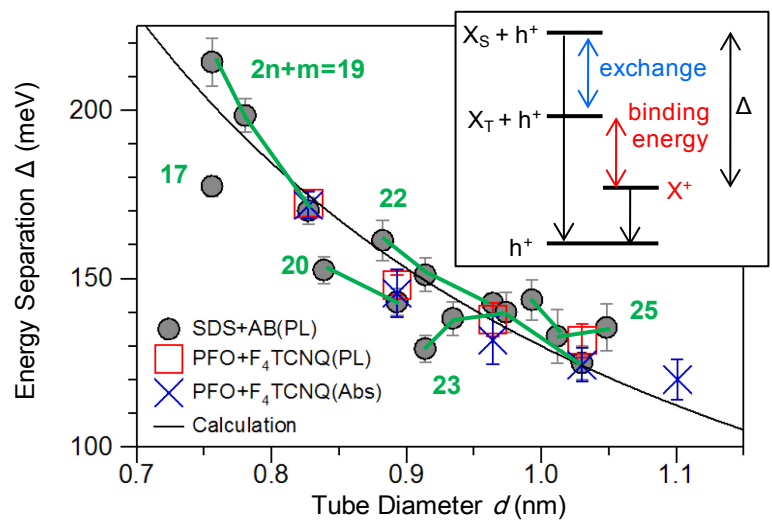

Fig. 4 Matsunaga et al.,